\documentclass[twocolumn,showpacs,preprintnumbers,amsmath,amssymb,prb,floatfix,superscriptaddress]{revtex4}

\usepackage{graphicx,tabularx}
\usepackage{amssymb}
\usepackage{dcolumn}
\usepackage{color}% to create grayscales for the ball and stick images
\usepackage[mathcal]{euscript}
\usepackage{color}
\definecolor{gray0}{gray}{0.0}%black
\definecolor{gray64}{gray}{0.25}
\definecolor{gray128}{gray}{0.5}
\definecolor{gray192}{gray}{0.75}
\definecolor{gray255}{gray}{1.0}%white

\def\CeIII{Ce$^{\mathrm{III}}$}
\def\LaIII{La$^{\mathrm{III}}$}
\def\CeIV{Ce$^{\mathrm{IV}}$}
\def\Ce3{Ce$^{\mathrm{III}}$}
\def\La3{La$^{\mathrm{III}}$}
\def\Ce4{Ce$^{\mathrm{IV}}$}
\def\A3{A$^{\mathrm{III}}$}
\def\B4{B$^{\mathrm{IV}}$}

\def\LZO{La$_2$Zr$_2$O$_7$}
\def\LCO{La$_2$Ce$_2$O$_7$}
\def\LCOg{La$_x$Ce$_{1-x}$O$_{2-x/2}$}

\def\CO2{CeO$_2$}
\def\C2O3{Ce$_2$O$_3$}

\def\Evf{E$_{vac,f}$}

\begin{document}

\title{DFT study of \LCO: disordered fluorite vs pyrochlore structure}
\author{D. E. P. Vanpoucke}
\affiliation{Dept. Inorganic and Physical Chemistry, Ghent University, Krijgslaan $281$ - S$3$, $9000$ Gent, Belgium}
\author{P. Bultinck}
\affiliation{Dept. Inorganic and Physical Chemistry, Ghent University, Krijgslaan $281$ - S$3$, $9000$ Gent, Belgium}
\author{S. Cottenier}
\affiliation{Center for Molecular Modeling, Ghent University, Technologiepark $903$, $9053$ Zwijnaarde, Belgium}
\author{V. Van Speybroeck}
\affiliation{Center for Molecular Modeling, Ghent University, Technologiepark $903$, $9053$ Zwijnaarde, Belgium}
\author{I. Van Driessche}
\affiliation{Dept. Inorganic and Physical Chemistry, Ghent University, Krijgslaan $281$ - S$3$, $9000$ Gent, Belgium}

\date{\today}
\begin{abstract}
The crystal structure of Lanthanum Cerium Oxide (\LCO) is investigated using \textit{ab initio} density functional theory (DFT) calculations. The relative stability of fluorite- and pyrochlore-like structures is studied through comparison of their formation energies. These formation energies show the pyrochlore structure to be favored over the fluorite structure, apparently contradicting the conclusions based on experimental neutron and X-ray diffraction (XRD).
By calculating and comparing XRD spectra for a set of differently ordered and random structures, we show that the pyrochlore structure is consistent with diffraction experiments. For these reasons, we suggest the pyrochlore structure as the ground state crystal structure for \LCO.
%we show that among the structures considered in this work, the pyrochlore geometry is clearly favorable over the disordered fluorite geometry.
\end{abstract}

\pacs{  } %--> wtf???
%----------------------------------------------------------------
\maketitle
%------------------------------------------------------------------------------------------
%------------------------------Introduction------------------------------------------------
%------------------------------------------------------------------------------------------
\section{Introduction}
\indent Since the pioneering work of Zintl and Croatto,\cite{ZintlE:ZAAC1939} \LCO\ has been studied for over half a century. During this time, \LCO\ and more generally the \LCOg\ compounds have been studied in the context of three-way automotive catalysts,\cite{DeganelloF:SSI2002,DeganelloF:SSI2003,ReddyB:2010CM,LiangShuang:JMaterSci2011} as an ionic conductor in solid oxide fuel cells,\cite{TullerHL:JES1975,KudoT:JES1975,BaeJongSung:JoECS2004} and as oxygen sensor.
In recent years, it has also become of interest as a new material for thermal barrier coatings.\cite{CaoXueqiang:AdvMat2003,MaWen:ScrMat2006,CaoXueqiang:JMatSciTech2007}
 %due to its low thermal conductivity, large thermal expansion coefficient and high phase stability even at high temperature,
It might also be considered as a new buffer layer in combination with perovskite superconductors for the use in coated conductors on Rolling Assisted Biaxially Textured substrates (RABiTS).\cite{VanDriesscheI:PAChem2002,VanDriesscheI:THERMEC2003,VanDriesscheI:2002ECVII}\\
\indent Although this compound has been known for a long time, its crystal structure remains a point of discussion. The two competing models for its crystal structure are the disordered fluorite and the pyrochlore structures.\\
\indent Many ternary oxides with the formula A$_2$B$_2$O$_7$, with $+$III ions A and $+$IV ions B, adopt a pyrochlore structure, making the latter a good candidate for the \LCO\ structure. Conversely, in many cases where a pyrochlore crystal structure is observed, the A and B ions are indeed lanthanides and/or transition metals. A pyrochlore structure (space group $Fd\bar{3}m$) can be obtained from a fluorite structure (space group $Fm\bar{3}m$) with one eighth of the oxygen ions missing ($Fd\bar{3}m$ Wickoff $8a$ site, when placing the origin at a B cation). Each oxygen vacancy is surrounded by four \B4\ ions (16$c$ sites), while six oxygen ions ($48f$ sites) are each surrounded by two \A3\ and two \B4\ ions, and the seventh oxygen ion ($8b$ site) is surrounded by four \A3\ ions ($16d$ sites). The pyrochlore structure has a clear short-range order. It turns out to be stable only in a certain range of ionic radius ratios of the two cations. Outside this range, a disordered fluorite structure becomes more stable.\cite{BrisseF:CJChem1967}\\
\indent This \emph{disordered} fluorite  structure for \LCO\ is obtained by random replacement of half the Ce cations in the CeO$_2$ cubic fluorite lattice by La cations. In addition, one eighth of the O anions are removed, also through random selection. This structure has the required stoichiometry and the crystal structure is a cubic fluorite with each cation site occupied by $0.5$ Ce and $0.5$ La atoms on average, and each anion site on average occupied by $0.875$ O atoms. As a result, no short-range order is present in a disordered fluorite structure, in contrast to the pyrochlore structure, .\\
\indent The border between the stability regions of the disordered fluorite and the pyrochlore structures has been defined empirically, to reflect the experimental data as well as possible.\cite{MinerviniL:JAmCeramSoc2000} With the ionic radius ratio for \LCO\ located very close to this border, \LCO\ itself becomes interesting for investigating the order-disorder transition in pyrochlores. Minervini \textit{et al.} show in their theoretical study of disorder in pyrochlore oxides that \LCO\ lies at the boundary of stability for pyrochlore formation. They claim that in this boundary region \LCO\ remains a disordered fluorite structure, though the pyrochlore structure appears stable with regard to cation and anion disorder.\cite{MinerviniL:JAmCeramSoc2000}\\
\begin{figure*}[!tb]
%\includegraphics[width=13cm,keepaspectratio=true]{images/LCO_Config_Novac_CeO2basis.png}\\
  \includegraphics[width=13cm,keepaspectratio=true]{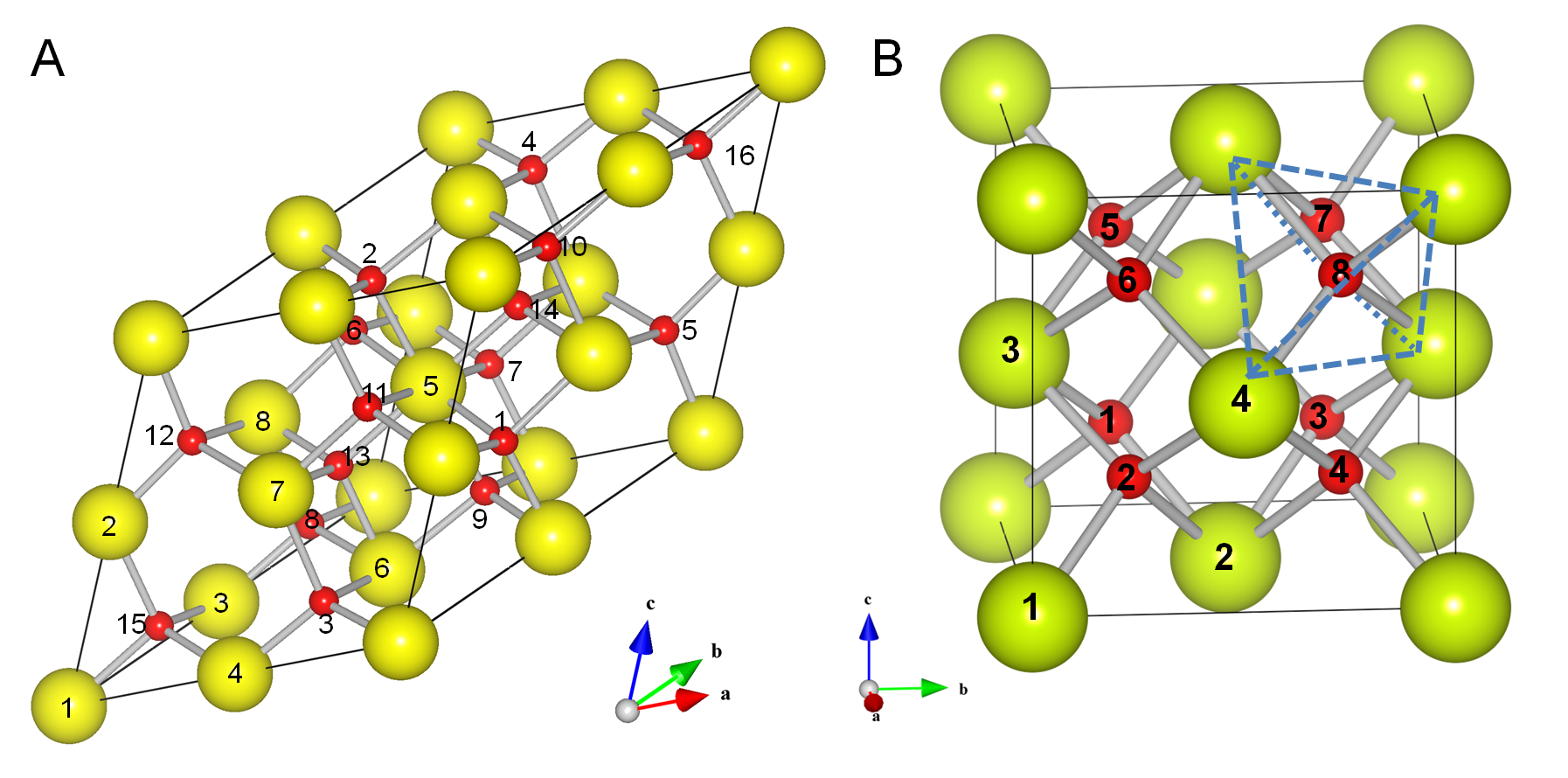}\\
  \caption{(Color online) Ball and stick representations of (A) the fcc primitive $2\times 2\times 2$ (p222) and (B) the cubic (c111) \CO2\ fluorite super cells. All the inequivalent Ce (big yellow spheres) and O (small red spheres) atom positions are indicated. The dashed tetrahedron on the cubic cell indicates the surrounding tetrahedron around the oxygen atom $8$.}\label{fig:LCO_config_Novac}
\end{figure*}
\indent Furthermore, there is a list of experiments which are in favor of the disordered fluorite structure. In their early neutron diffraction experiments, Brisse and Knop claim \LCO\ to have a disordered fluorite structure rather than a pyrochlore one.\cite{BrisseF:CJChem1967} But also more recent XRD experiments point towards a disordered fluorite structure, due to the lack of pyrochlore-specific peaks in the spectrum.\cite{YamamuraH:JCSJ2003,YamamuraH:JCSJ2004} It was also shown that this structure remains stable, even at very high temperatures.\cite{CaoXueqiang:AdvMat2003,MaWen:ScrMat2006}\\
\indent In contrast, recent studies of \LCOg\ show a different behavior.
XRD experiments by Bae \textit{et al.}\cite{BaeJongSung:JoECS2004} show
that the CeO$_2$ fluorite structure is maintained for La concentrations up to $x=0.40$ (for \LCO\ $x=0.50$), and only at higher La concentrations the pyrochloric ordering of the cations (\LaIII\ and \CeIV) appears. These authors also observe that the lattice parameter does not expand in a linear fashion as was reported earlier by Chambonnet \textit{et al.}\cite{ChambonnetD:PC1998}, and assume this reduced expansion is due to the clustering of O-vacancies. Other authors find multiple non-linear regions in the expansion of the lattice parameter. Ryan \textit{et al.}\cite{RyanKM:JPCondMat2003} link this behavior to the presence of two phases with different La concentration. Of these, the phase with the high La density shows a discontinuity in the lattice parameter just around $38$\% La, and a constant lattice parameter above $40$\% La.
Though they do report observing local ordering in the system, unlike Bae \textit{et al.}\cite{BaeJongSung:JoECS2004}, they refrain from linking the second phase to a pyrochlore phase. In contrast, O'Neill and Morris find only a single phase in their investigation of \LCOg.\cite{ONeillWM:ChemPhysL1999} However, they only go up to a La concentration of 10\%, where Ryan \textit{et al.}\cite{RyanKM:JPCondMat2003} observed the appearance of the second phase at a La concentration of $\sim20$\%, indicating that below this La concentration, the second phase was either not present or not distinguishable.\\
\indent As a result of this complex behavior, no definitive crystal structure has been established for \LCO, and there exists only a single entry in the ICSD database, referring back to the work of Zintl and Croatto.\cite{ZintlE:ZAAC1939,ICSD1,ICSD2}
Experiments are interpreted in favor of the disordered fluorite as well as the pyrochlore structure, and some authors even refer to \LCO\ both as a pyrochlore and a fluorite structure in the same work.\cite{LopesFWB:Hydrometallurgy2009}\\
% Although the possible ionic sites for both structures are the same, the latter structure is highly ordered, while for the former the site occupation is assumed to be random and without local order.\\
\indent In this paper we present an \textit{ab initio} density functional theory (DFT) study of the \LCO\ crystal structure. Based on the order--disorder contrast of the experimentally proposed pyrochlore and disordered fluorite structures, we focus on the effect of order in possible \LCO\ structures. By calculating and comparing the formation energies and XRD spectra for a set of different ordered and random structures, we show the pyrochlore structure to be favored over the disordered fluorite structure.
%%%%%%%%%%%%%%%%%%%%%%%%%%%%%%%%%%%%%%%%%%%%%%%%%%%%%%%%%%%%%%%%%%%%%%%%%%%%%%%%%%%%%%%%%%
\begin{table}[!tb]
\caption{O-vacancy and La positions for the different structures investigated. The tetrahedral surrounding of the O-vacancy is indicated as $x$Ce$y$La with $x$ and $y$ the number of Ce and La atoms in the tetrahedron surrounding the vacancy. The notation NV indicates no vacancies are present in the system. The indexes are shown in Fig.~\ref{fig:LCO_config_Novac}A and B. The LZO and L$111$ structures are generated using the (fluorite) p$222$ unit cell (\textit{cf.} Fig.~\ref{fig:LCO_config_Novac}A).}\label{table:LCOconfig}
\begin{ruledtabular}
\begin{tabular}{l|ccc}
& La positions & O-vacancies & space group \\
\hline
c111 L$1_0$ NV & 1, 2& $-$    & P4/mmm \\
c111 L$1_0$ 2Ce2La & 1, 2 & 2 & Cmm2   \\
\hline
LZO NV & 5, 6, 7, 8 &  $-$    & Fd$\bar{3}$m \\
LZO 2Ce2La & 5, 6, 7, 8 & 1, 6& C2/m \\
LZO 4Ce & 5, 6, 7, 8 & 15, 16 & Fd$\bar{3}$m \\
LZO 4La & 5, 6, 7, 8 & 13, 14 & Fd$\bar{3}$m \\
%\hline
%p222 L$1_0$ NV & 1, 2, 5, 6 & $-$\\
%p222 L$1_0$ 2Ce2La & 1, 2, 5, 6 & 1, 6 \\
\hline
L111 NV & 1, 6, 7, 8 &  $-$   & R$\bar{3}$m\\
L111 3Ce1La & 1, 6, 7, 8 & 15, 16& R$\bar{3}$m\\
%L111 3Ce1La b& 1, 6, 7, 8 & 1, 7 \\
%L111 3Ce1La c& 1, 6, 7, 8 & 7, 9 \\
L111 1Ce3La & 1, 6, 7, 8 & 13, 14& R$\bar{3}$m\\
\end{tabular}
\end{ruledtabular}
\end{table}
\begin{figure}[tb!]
%\includegraphics[width=8cm,keepaspectratio=true]{images/LCO_Config_Novac_cations.png}\\
  \includegraphics[width=8cm,keepaspectratio=true]{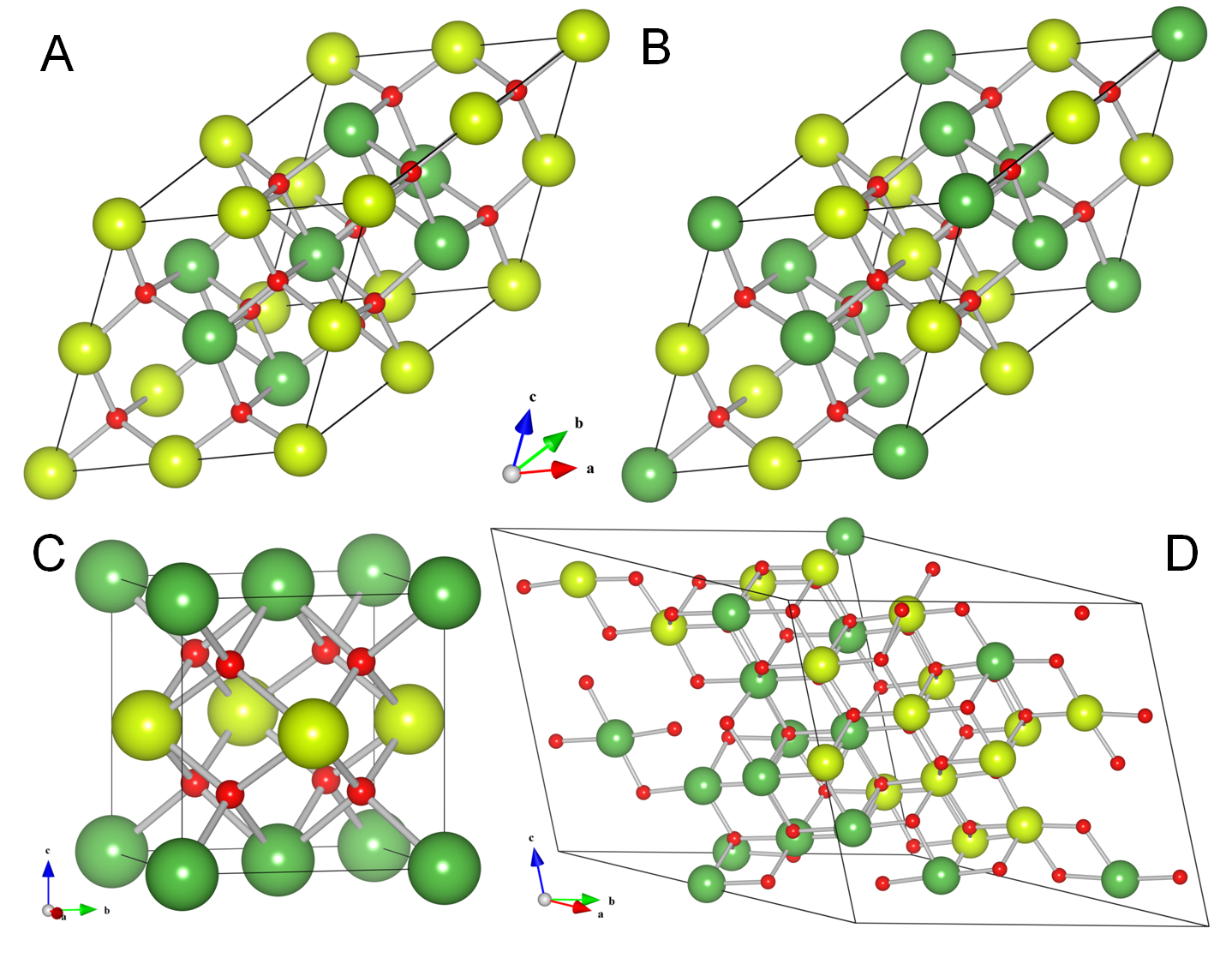}\\
  \caption{(Color online) Ball and stick representations of the different cation distributions, as constructed from Table~\ref{table:LCOconfig} and Fig.~\ref{fig:LCO_config_Novac} or taken from literature: (A) the LZO, (B) the L$111$, (C) the c$111$ L$1_0$ and, (D) the SQS distribution.\cite{JiangC:PRB2009} Red, green and yellow spheres show the O, La and Ce positions, respectively.}\label{fig:LCO_geom}
\end{figure}
\section{Theoretical method}
\indent Starting from the \CO2-fluorite structure, the \LCO\ structure is obtained by replacing half of the \Ce4 ions by \La3 (\textit{i.e.} CeO$_2$ $\rightarrow$ La$_2$Ce$_2$O$_8$), and, due to charge compensation, removing one eight of the O ions (\textit{i.e.} La$_2$Ce$_2$O$_8$ $\rightarrow$ La$_2$Ce$_2$O$_7$). This can be achieved in a large number of ways, resulting in both ordered structures and disordered structures. Because an exhaustive study is impossible within the DFT framework, and because the most prominent difference between the disordered fluorite and pyrochlore structures is their amount of short range order, our study can be confined to some highly ordered structures and a ``random structure''.
%In the following, ordered structures will refer to short range order of the cations.
This allows us to investigate the influence of cation-ordering on the stability of the system. Moreover, the chemical environment of the vacancies can be studied in this way.\\
\indent The three ordered cation distributions we use are: the \LZO\ pyrochlore structure (LZO), the cubic CuAu L$1_0$ structure (c$111$ L$1_0$), with alternating La and Ce layers along the [001] orientation, and the L111 structure, with alternating Ce and La layers along the [111] orientation.
Both the LZO and L111 structures can be constructed using the fluorite p$222$ primitive fcc supercell, shown in Fig.~\ref{fig:LCO_config_Novac}A, the L$1_0$ structure can be constructed using the cubic fluorite supercell, shown in Fig.~\ref{fig:LCO_config_Novac}B. For each of these structures a set of O-vacancies is chosen in such a way that all vacancies in a single system have the same chemical environment.
Table \ref{table:LCOconfig} gives the positions of the La cations and O-vacancies in the unit cells used for these structures, and shown in Fig. \ref{fig:LCO_config_Novac}A and B, and the resulting cation distributions are shown in Fig.~\ref{fig:LCO_geom}A, B, and C. The tetrahedral surrounding (\textit{cf.}~Fig.~\ref{fig:LCO_config_Novac}B) of the vacancies is indicated as $x$Ce$y$La, with $x$ and $y$ integers giving the number of Ce and La cations in the tetrahedron. Cells without O-vacancies are indicated with NV.\\
%The L$1_0$ structure is generated with a cubic (c111) unit cell and a primitive (p222) unit cell. Though the NV versions are indistinguishable, the systems containing O-vacancies differ in their inter-vacancy distance.
\indent All p222 cells, except the NV structures, contain two vacancies. To improve the comparability of the different structures, the two O-vacancies are always placed symmetrically around a single cation. This way we can assume the contribution to the formation energy due to the O-vacancy interaction energy to be the same for all p222 systems.
% The two exceptions, L111 3Ce1La b and c, are added to study the O-vacancy interaction in detail.
As a consequence of this setup, all four cations surrounding the O-vacancy in the p222 cells are $6$-coordinated. Note that we use the term coordination only to indicate the number of anions present surrounding the cation, not the number of anions the cation has an actual bond with.\\
\indent There are several methods in literature to obtain physical properties of disordered or random structures. The most intuitive perhaps is the use of averaged potentials in the Virtual Crystal Approximation, where instead of using a specific potential for each ionic species one uses a potential representing the average atomic species present.\cite{RamerNJ:PRB2000}
% Though this method has shown to provide reasonable agreement
%\textbf{some pro \& cons: not available in vasp, good agreement with exp, but some prop can not be modeled because env.specific.}
Another approach is the use of Monte-Carlo methods, and a statistical analysis of the results obtained. However, this method requires one to do thousands (or even more) of calculations, which is not feasible due to the size of our unit cell and the cost of DFT calculations. A third possible approach is the use of cluster expansions, where the configurational dependence of the physical property of interest is described as a sum over bonds or clusters. However, the configurational dependence of any variable can be described exactly by a cluster expansion only if all terms are retained, which is not the case in any practical implementation. Furthermore, predictions of the cluster expansion can depend sensitively on the choices made by the user with regard to for example input structures.\cite{LerchD:ModSimMSE2009,SanchezJM:PhysA11984}
We will make use of what is called a Special Quasi-random Structure (SQS). \cite{Zunger_SQS:PRL1990,Zunger_SQS:PRB1990,Ruban:PRB2003} This is a super cell chosen in such a way that the correlation functions of the system resemble those of a truly random system as closely as possible. Although the super cell is a few times larger than a unit cell, and has no symmetry, the fact that we only need to do calculations with a single structure makes this technique orders of magnitude cheaper than a Monte-Carlo method. Neither is it necessary to use statistical methods to obtain system properties.\\
\indent The main problem with using SQS is the fact that they need to be constructed first. For the pyrochlore(fluorite) system an SQS containing $88(96)$ atoms was previously given by Jiang \textit{et al.}\cite{JiangC:PRB2009}. We use this SQS to investigate the physical properties of a disordered fluorite structure. A ball and stick representation is shown in Fig. \ref{fig:LCO_geom}D. Because all possible tetrahedral surroundings for the vacancies are present in this system, no tetrahedral surrounding is indicated.\\
%\begin{figure}[tb!]
%\includegraphics[width=8cm,keepaspectratio=true]{images/PBE_L2C2O7_SQS_Vac.png}\\
%\caption{Ball and stick representation of the \LCO-SQS super cell
%  used.\cite{JiangC:PRB2009} Red, green and yellow spheres indicate the O, La
%  and Ce ion positions, respectively.}\label{fig:PBE_LCO_SQSvac}
%\end{figure}
\indent The electronic structure calculations are performed within the DFT framework using the projector augmented wave (PAW) approach for the core-valence interaction and the Perdew-Burke-Ernzerhof (PBE) approximation for the exchange-correlation functional as implemented in the \textsc{VASP} code.\cite{Blochl:prb94,Kresse:prb99,Kresse:prb93,Kresse:prb96,PBE_1996prl}
The kinetic energy cutoff is set at $500$ eV and special $k$-point sets of
$8\times 8\times8$, $4\times 4\times4$, and $4\times 4\times4$ $k$-points are used for static self-consistent calculations in the c111, p222 and the SQS cells respectively. For the SQS cells a smaller $2\times 2\times2$ $k$-point set is used during relaxation. To optimize the geometry a conjugate gradient algorithm is applied. Both ion positions and cell parameters are optimized simultaneously.\\
\indent Because the $4f$ electron of Ce is known to be badly described in DFT at the PBE-level, we also performed calculations including on-site Coulomb corrections (PBE+U).\cite{Svane:PRB1996,Skorodumova:prb2001} The inclusion of on-site Coulomb corrections has been shown to provide a consistent treatment of Ce ions in ceria.\cite{Nolan:ss2005a,Fabris:prb2005,Andersson:prb2007,Loschen:prb2007,DaSilva:prb2007b}
The choice of U for the Ce $4f$ electrons is found to be optimal in the range of $2$--$8$ eV, depending on which property is under specific investigation.\cite{Andersson:prb2007,Loschen:prb2007,castleton:jcp2007}
For reduced ceria, and a consistent description of pure \CO2\ and \C2O3, both Nolan \textit{et al.}\cite{Nolan:ss2005a} and Andersson \textit{et al.}\cite{Andersson:prb2007} suggest for GGA+U a U$_{\mathrm{Ce}}= 5.0$ eV. Because \LCO\ contains a large amount of O vacancies, we will follow this suggestion.\\
%, and use U$_{\mathrm{Ce}}=5.0$ eV.\\
\begin{table*}[!tb]
\caption{The heat of formation $\Delta$H$_f$ and vacancy formation energy \Evf\  as calculated using equations \eqref{eq:LCO_HeatofForm} and \eqref{eq:LCO_VacFormE}. The inter-vacancy distance d$_{\mathrm{vac}-\mathrm{vac}}$ is calculated as the distance between the centers of mass of the cation tetrahedra surrounding the O-vacancies. Due to the random nature of the SQS system no inter-vacancy distance is given. The notation NV is used to identify systems without vacancies. For the PBE+U calculations we use U$_{\mathrm{O}}=7.0$ eV on the O $2p$ states and U$_{\mathrm{Ce}}=5.0$ eV on the Ce $4f$ states for all calculations involved.}\label{table:EnergyLCO}
\begin{ruledtabular}
\begin{tabular}{l|cccccc}
& \multicolumn{3}{c}{PBE} &  \multicolumn{3}{c}{PBE+U} \\
\cline{2-4} \cline{5-7}
& $\Delta$H$_f$ & \Evf & d$_{\mathrm{vac}-\mathrm{vac}}$  & $\Delta$H$_f$ & \Evf & d$_{\mathrm{vac}-\mathrm{vac}}$\\
&  (eV)&  (eV)& (\AA)& (eV)&  (eV)& (\AA)\\
\hline
c111 L$1_0$ NV    & 1.177 & $-$    &  $-$   & 1.280 & $-$ & $-$   \\
c111 L$1_0$ 2Ce2La& 0.758 & -0.419 &  $5.576$ & 0.800 & -0.480 & $5.570$ \\
%p222 L$1_0$ NV & 1.178 & $-$ & 1.281 & $-$ \\
%p222 L$1_0$ 2Ce2La& 1.007 & -0.171 & 1.115 & -0.166 \\
\hline
LZO NV & 1.167 & $-$ & $-$ & 1.364 & $-$ & $-$  \\
LZO 2Ce2La& 0.939  & -0.229 & $4.829$ &  1.105 & -0.259 & $4.818$\\
LZO 4Ce   & -0.016 & -1.183 & $4.861$ &  0.220 & -1.143 & $4.862$ \\
LZO 4La   & 1.668  &  0.501 & $4.812$ &  1.620 &  0.256 & $4.803$ \\
\hline
L111 NV  & 1.169 & $-$ & $-$ & 1.564 & $-$ & $-$ \\
L111 3Ce1La  & 0.452 & -0.717 & $4.782$ & 0.668 & -0.896 & $4.787$ \\
%L111 3Ce1La b & 0.072 & -1.097 & 0.198 & -1.366 \\
%L111 3Ce1La c & 0.123 & -1.046 & 0.263 & -1.301 \\
L111 1Ce3La & 1.352 & 0.183 & $4.843$ & 1.420 & -0.144 & $4.843$ \\
\hline
SQS NV& 1.160 & $-$ & $-$ & 0.699 & $-$ & $-$ \\
SQS & 0.527 & -0.633 & n/a & 0.607 & -0.91 & n/a \\
\end{tabular}
\end{ruledtabular}
\end{table*}
\indent In theoretical studies of La doping of ceria surfaces, it was also shown by Yeriskin and Nolan that an on-site Coulomb correction is required to correctly describe the localized hole on the O $2p$ state.\cite{YeriskinI:JPCondMat2010} This localized hole results from the introduced \LaIII\ ion at a \CeIV\ site. Since the \LCO\ structure can be interpreted as an extreme case of La doping we use the same U$_{\mathrm{O}}=7.0$ eV on the O $2p$ states.\\
\indent The XRD spectra are generated for the relaxed structures using the visualizer tool of the ICSD database.\cite{ICSD1,ICSD2}
%%%%%%%%%%%%%%%%%%%%%%%%%%%%%%%%%%%%%%%%%%%%%%%%%%%%%%%%%%%%%%%%%%%%%%%%%%%%%%
%%%%%%%%%%%%%%%%%%%%%%%%%%%%_RESULTS_%%%%%%%%%%%%%%%%%%%%%%%%%%%%%%%%%%%%%%%%%
%%%%%%%%%%%%%%%%%%%%%%%%%%%%%%%%%%%%%%%%%%%%%%%%%%%%%%%%%%%%%%%%%%%%%%%%%%%%%%
\section{Results and Discussion}
\subsection{Energetics}
\indent The energetics of the system are studied through comparison of the heat of formation $\Delta$H$_f$ and the vacancy formation energies \Evf\ of the different configurations. The heat of formation is calculated using:
\begin{equation}\label{eq:LCO_HeatofForm}
\Delta\mathrm{H}_f=\big(E_{tot}-N_{Ce}E_{CeO_2}-\frac{1}{2}N_{La}E_{La_2O_3}-\frac{1}{2}N_{O_{add}}E_{O_2}\big)/N_f,
\end{equation}
with $E_{tot}$ the total ground state energy of the relaxed system, $N_{Ce}$ and $N_{La}$ the number of Ce and La atoms present in the system, $E_{CeO_2}$ and $E_{La_2O_3}$ the bulk energy of CeO$_2$ and La$_2$O$_3$, $N_{O_{add}}$ the number of additional O atoms not accounted for by the CeO$_2$ and La$_2$O$_3$ formula units, and $E_{O_2}$ the ground state energy calculated for a free oxygen molecule. The factor $N_f$ is the number of formula units, which we define as the size of a supercell containing a single O-vacancy. The equivalent supercell is used for the systems without vacancies.\\
\indent The O-vacancy formation energy \Evf\ is calculated using:
\begin{equation}\label{eq:LCO_VacFormE}
\mathrm{E}_{vac,f} = \big(E_{\mathrm{vac}}+\frac{1}{2}N_{\mathrm{vac}}E_{O_2}-E_{\mathrm{novac}}\big)/N_{\mathrm{vac}},
\end{equation}
with $E_{\mathrm{vac}}$ and $E_{\mathrm{novac}}$ the ground state energies obtained for the systems with and without vacancies, respectively, and $N_{\mathrm{vac}}$ the total number of vacancies present in the system with O-vacancies. Negative values indicate a stabilization of the system. The resulting energies are shown in Table \ref{table:EnergyLCO}. \\
\indent The calculated heats of formation and O-vacancy formation energies presented in Table~\ref{table:EnergyLCO} show some clear trends with regard to the stability of \LCO. If no O-vacancies are involved (\textit{i.e.}~La$_2$Ce$_2$O$_8$) PBE calculations show all possible cation distributions to be roughly degenerate. Including on site Coulomb corrections for Ce and O lifts this degeneracy and shows the disordered fluorite structure to be the most stable. However, \LCO\ contains $12.5$\% O-vacancies compared to the \CO2\ fluorite structure to provide the necessary charge compensation for the introduction of trivalent La in a tetravalent Ce position in the \CO2\ lattice. In contrast to the previous, the introduction of O-vacancies shows no qualitative difference between the PBE and PBE+U results. In both cases the introduction of O-vacancies is beneficial for the heat of formation. The sole exceptions are the systems with three or four La cations in the tetrahedral surrounding of the O-vacancy: L$111$ $1$Ce$3$La (PBE only) and LZO $4$La (PBE and PBE+U).\\ %Note that without the Coulomb correction, the configurations without O-vacancies have a nearly degenerate heat of formation. This degeneracy is clearly lifted in the PBE+U case. The SQS is in both cases energetically the most favorable system. In contrast, the introduction of vacancies shows no qualitative difference between the PBE and PBE+U results. In both cases the introduction of O-vacancies is beneficial for the heat of formation. The sole exceptions are the systems with three or four La cations in the tetrahedral surrounding of the O-vacancy: L$111$ $1$Ce$3$La (PBE only) and LZO $4$La (PBE and PBE+U).\\
\indent Of all configurations studied, the LZO 4Ce system (\textit{i.e.} the actual pyrochlore structure) has the most favored heat of formation and O-vacancy formation energy. It has a heat of formation that is roughly $0.5$ eV/formula unit better than the SQS system, representing the disordered fluorite structure. %In addition, the O-vacancy concentration necessary to have an \LCO-system makes it impossible for the LZO 4Ce system to have a different ionic configuration.
Closer examination of \Evf\ with regard to the tetrahedral surrounding of the O-vacancy shows the following order of increasing stability:
\[
4\mathrm{La} < 1\mathrm{Ce}3\mathrm{La} < 2\mathrm{Ce}2\mathrm{La} < 3\mathrm{Ce}1\mathrm{La} < 4\mathrm{Ce}.
\]
This shows there is a clear correlation between the number of Ce ions in the surrounding tetrahedron and \Evf. This result is independent of the introduced Coulomb correction, and results for both PBE and PBE+U functionals in an improvement of \Evf\ with $\sim1.5$ eV, going from the least to the most stable configuration. This also means that ordered structures containing Ce tetrahedra show large stability when the O-vacancies are enclosed in these Ce tetrahedra. On the other hand, structures with no Ce in the tetrahedron surrounding the O-vacancy show poor stability (Compare the LZO 4Ce and LZO 4La structures in Table~\ref{table:EnergyLCO}. In both cases the cations have a pyrochlore geometry). Due to this stability trend for O-vacancy positions, it is clear why the disordered fluorite structure (\textit{i.e.} SQS) is less favorable than ordered structures. Because of the random nature of the disordered fluorite structure, fewer pure Ce tetrahedra are present. In addition, because of their random distribution, the O-vacancies are also placed in less favorable surroundings than could be the case in ordered structures.\\
%\indent Although there is a clear correlation between the energetics and the number of Ce ions in the vacancy surrounding,
\indent Irrespective of the correlation between \Evf\ and the number of Ce cations in the tetrahedral O-vacancy surrounding, there still appears to be quite some variation in formation energies for the same surrounding when comparing the two $2$Ce$2$La structures in Table \ref{table:EnergyLCO}. These two systems differ in two aspects: the inter-vacancy distance and the coordination of the cations in the system. This leads to the question which of both is responsible for the difference in vacancy formation energy.\\
%This leads to the question wether the difference in vacancy formation energy is due to the difference in distance between the two vacancies, or due to the change in coordination of the Ce atoms.\\
\begin{table}[!tb]
\caption{O-vacancy and La positions for some additional structures. The `L111 3Ce1La' structure of Table \ref{table:LCOconfig} has been repeated here as `L111 3Ce1La A' for comparison.}\label{table:LCOconfig2}
\begin{ruledtabular}
\begin{tabular}{l|ccc}
& La positions & O-vacancies & space group\\
\hline
p222 L$1_0$ NV & 1, 2, 5, 6 & $-$ & P4/mmm\\
p222 L$1_0$ 2Ce2La & 1, 2, 5, 6 & 1, 6 &Imma\\
\hline
L111 3Ce1La A& 1, 6, 7, 8 & 15, 16& R$\bar{3}$m\\
L111 3Ce1La B& 1, 6, 7, 8 & 1, 7  & Pm\\
L111 3Ce1La C& 1, 6, 7, 8 & 7, 9  & C2/m\\
\end{tabular}
\end{ruledtabular}
\end{table}
\begin{table*}[!tb]
\caption{The same as Table~\ref{table:EnergyLCO}, for the additional structures. In addition, also the distribution of the Ce coordination is given. The c$111$ L$1_0$ $2$Ce$2$La and L$111$ $3$Ce$1$La A structures are repeated from Table~\ref{table:EnergyLCO} for comparison. }\label{table:EnergyLCO2}
\begin{ruledtabular}
\begin{tabular}{l|ccccccc}
&Ce-Coord. & \multicolumn{3}{c}{PBE} &  \multicolumn{3}{c}{PBE+U} \\
\cline{3-5} \cline{6-8}
& ($6|7|8$)-fold & $\Delta$H$_f$ & \Evf & d$_{\mathrm{vac}-\mathrm{vac}}$& $\Delta$H$_f$ & \Evf & d$_{\mathrm{vac}-\mathrm{vac}}$\\
& (\%) &  (eV)&  (eV)& (\AA)&  (eV)&  (eV)& (\AA)\\
\hline
p222 L$1_0$ NV & ($0|0|100$) &1.178 & $-$ & $-$ & 1.281 & $-$ & $-$ \\
p222 L$1_0$ 2Ce2La& ($50|0|50$) & 1.007 & -0.171 & $4.784$ & 1.115 & -0.166 & $4.776$ \\
c111 L$1_0$ 2Ce2La& ($0|100|0$) & 0.758 & -0.419 &  $5.576$ & 0.800 & -0.480 & $5.570$ \\
\hline
L111 3Ce1La A & ($75|0|25$) & 0.452 & -0.717 & $4.782$ & 0.668 & -0.896 & $4.787$ \\
L111 3Ce1La B & ($50|50|0$) & 0.072 & -1.097 & $3.960$ & 0.198 & -1.366 & $3.945$ \\
L111 3Ce1La C & ($50|50|0$) & 0.123 & -1.046 & $2.937$ & 0.263 & -1.301 & $2.934$ \\
\end{tabular}
\end{ruledtabular}
\end{table*}
\begin{table*}[!tb]
\caption{Geometric data for the PBE and PBE+U results. We use the notation NV to indicate systems without vacancies. For these NV systems $\Delta$V is calculated with regard to the volume of the equivalent CeO$_2$ fluorite cell, while for the other systems it is calculated with regard to the NV system. The lattice parameter $a$ (see text) is calculated as $a=\sqrt[3]{V_f}$,
and $\Delta$a gives the change of this lattice parameter with regard to the lattice parameter of CeO$_2$ fluorite and the calculated lattice parameter of the NV system.}\label{table:LCO_GeomPBE}
\begin{ruledtabular}
\begin{tabular}{l|cccccccc}
& \multicolumn{4}{c}{PBE} & \multicolumn{4}{c}{PBE+U} \\
\cline{2-5} \cline{6-9}
& $\Delta$V & a & \multicolumn{2}{c}{$\Delta$a} & $\Delta$V & a & \multicolumn{2}{c}{$\Delta$a} \\
&   &   &  vs CeO$_2$ & vs NV &   &   &  vs CeO$_2$ & vs NV \\
&  (\%)&  (\AA)&  (\%)&  (\%)&  (\%)&  (\AA)&  (\%)&  (\%)\\
\hline
c111 L$1_0$ NV  & 8.044  & 5.606 & 2.613 & $-$ & 7.945 & 5.609 & 2.581 & $-$ \\
c111 L$1_0$ 2Ce2La & -1.260 & 5.582 & 2.180 & -0.422 & -1.631 & 5.579 & 2.021 & -0.547 \\
p222 L$1_0$ NV  & 8.080  & 5.606 & 2.624 & $-$ & 7.947 & 5.609 & 2.582 & $-$  \\
p222 L$1_0$ 2Ce2La & -1.281 & 5.582 & 2.184 & -0.429 & -1.725 & 5.577 & 1.989 & -0.578  \\
\hline
LZO NV  & 8.328 & 5.611 & 2.702 & $-$   & 8.479 & 5.618 & 2.750 & $-$ \\
LZO 2Ce2La & -0.668 & 5.598 & 2.473 & -0.223 & -1.533 & 5.590 & 2.222 & -0.514 \\
LZO 4Ce    & 0.108 & 5.613 & 2.739 & 0.036 & -0.228 & 5.614 & 2.672 & -0.076\\
LZO 4La    & -2.889 & 5.556 & 1.704 & -0.973 & -3.793 & 5.547 & 1.434 & -1.281 \\
\hline
L111 NV   & 8.253 & 5.609 & 2.679 & $-$ & 8.639 & 5.621 & 2.800 & $-$  \\
L111 3Ce1La A & -0.292 & 5.604 & 2.579 & -0.097 & -1.045 & 5.602 & 2.441 & -0.350 \\
L111 3Ce1La B & -0.081 & 5.608 & 2.651 & -0.027& -0.889 & 5.605 & 2.495 & -0.297  \\
L111 3Ce1La C & 0.299 & 5.615 & 2.781 & 0.099 & -0.459 & 5.613 & 2.643 & -0.153 \\
L111 1Ce3La  & -1.736 & 5.577 & 2.081 & -0.582& -2.774 & 5.569 & 1.841 & -0.933  \\
\hline
SQS NV & 8.530 & 5.614 & 2.766 & $-$ & 9.173 & 5.630 & 2.969 & $-$  \\
SQS   & -0.020 & 5.614 & 2.759 & -0.007 & -1.285 & 5.606 & 2.526 & -0.430 \\
\end{tabular}
\end{ruledtabular}
\end{table*}
\indent We have studied two sets of additional systems, given in Table~\ref{table:LCOconfig2}, to further investigate the influence of these aspects. The first set consists of the p$222$ L$1_0$ structure. This is the L$1_0$ distribution of the cations but this time constructed using a p222 unit cell (\textit{cf.} Fig.~\ref{fig:LCO_config_Novac} and Table~\ref{table:LCOconfig2}). The O-vacancies are placed at opposing sides of a single Ce cation,\cite{fn:oppsides} reducing the inter-vacancy distance by about $0.8$ \AA\ compared to the c111 L$1_0$ structure. At the same time, the coordination changes from 7-fold coordination for all cations in the c111 L$1_0$ structure to 6-fold coordination for half the Ce and half the La ions and 8-fold coordination for the remaining cations in the p222 L$1_0$ structure. %\cite{fn:coordination}\\
A second set of structures consists of L$111$ $3$Ce$1$La B and C (\textit{cf.} Table~\ref{table:LCOconfig2}), where the former L$111$ $3$Ce$1$La structure will be referred to as L$111$ $3$Ce$1$La A. Table~\ref{table:EnergyLCO2} gives the energies, inter-vacancy distances and Ce-coordination for these systems.\\ %Note that we use the term coordination only to indicate the number of anions present surrounding the cation, not the number of anions the cation has an actual bond with.\\
%The inter-vacancy distances are approximately $5.1$, $4.0$, and $2.6$ \AA\, for the L111 3Ce1La A, B, and C structures, respectively. In the L111 3Ce1La A structure three quarters of the Ce ions are 6-fold coordinate, and one quarter 8-fold coordinate, while for both the B and C structures half the Ce ions are 6-fold coordinate and half are 7-fold coordinate. The heat of formation and vacancy formation energy of all these additional systems is shown in Table \ref{table:EnergyLCO2}.\\
\indent Comparing L$1_0$ structures in Tables~\ref{table:EnergyLCO}~and~\ref{table:EnergyLCO2} shows that the heat of formation, $\Delta$H$_f$, is exactly the same, as it should be, when no vacancies are present. However, with O-vacancies included, there is a significant difference in \Evf\ of $0.2-0.3$ eV. This difference between the c$111$ and p$222$ L$1_0$ $2$Ce$2$La systems is related to their different vacancy configurations. %The two configurations differ in the coordination of the Ce cations present, but there is also a difference in inter-vacancy distance of $\sim0.8$ \AA.
Comparing the three $2$Ce$2$La systems studied, we find that both systems with half the Ce ions 6-fold coordinate are $\sim0.2$ eV less stable than the third system (c111 L$1_0$ 2Ce2La), without 6-fold coordinate Ce.\\
%This leads to the question if the difference in vacancy formation energy is
%due to the difference in distance between the two vacancies, or due to the
%change in coordination of the Ce atoms.\\
% here we investigate the effect of those configurations,
% is the distance an important factor or is it the coordination after all?
% If it is the distance, then this is an argument against clustering of
% vacancies.
\indent The three L$111$ $3$Ce$1$La configurations can also be used to distinguish the effects of distance and coordination. Due to the three different inter-vacancy distances and the same Ce coordination for the B and C structures one can consider following scenarios: If a large O-vacancy distance is beneficial to the system then the O formation energies should be ranked A$<$B$<$C, with A the best configuration. Contrary, if the opposite is the case, namely clustering or coalescing of the vacancies is beneficial, the C case should be the best structure, giving rise to a ranking C$<$B$<$A. In both these scenarios the Ce coordination is assumed to have little or no influence.
If on the other hand the inter-vacancy distance has little or no influence and the reduced number of $6$-fold coordinate Ce cations causes the decreased vacancy formation energy, then the B and C case should be (nearly) degenerate and more stable than the A case. The resulting ranking should then be A$>$B$=$C.\\
\indent Comparing the energies in Table \ref{table:EnergyLCO} and \ref{table:EnergyLCO2} we see a significant improvement ($\sim0.4$ eV) in the heat of formation and O-vacancy formation energy for the B and C configurations compared to A. This change is comparable to the one found for the L$1_0$ 2Ce2La structures. Table~\ref{table:EnergyLCO2} also shows the B and C configurations to be roughly degenerate. This shows that the improved energetics originate from the third scenario described above. The reduced number of $6$-fold coordinate Ce ions causes a significant improvement of the formation energy \Evf. Upon closer examination we also notice that the B configuration is slightly more stable ($\sim0.06$ eV) than the C configuration, showing that also the O-vacancy distance plays a role, albeit a minor one, and, more importantly, that the vacancies repel each other.\\
%\indent Furthermore, comparing the energetics of these configurations with the %LZO 4Ce configurations shows them to be nearly degenerate in the PBE case and %slightly more stable in the PBE+U case.\\
%then something about order of influence, tetrahedron config/coordination/vac-vac dist

\indent The results of this section also give us some additional understanding on the structure of \LCOg. The above calculations show that for \LCO\ the Ce atoms prefer to cluster in tetrahedra around the O-vacancies, or vice versa, the O-vacancies prefer to appear inside Ce tetrahedra. If we assume that this behavior is also valid for \LCOg, then the transformation of \CO2\ to \LCO\ can be understood as follows: For low La concentrations, and thus low O-vacancy concentrations, the vacancies, imbedded in Ce tetrahedra, will be distributed homogeneously throughout the \CO2\ fluorite structure. With increasing La concentration the number of available Ce cations for Ce tetrahedra reduces while the number of O-vacancies increases. For a system where the cations are randomly distributed this means that at La concentrations of $>42.8$\% there are too few Ce-tetrahedra available to accommodate all the O-vacancies. This seems to coincide with the experimentally observed discontinuity in the lattice parameter observed by Bae \textit{et al.}\cite{BaeJongSung:JoECS2004} %N(1-x)^4-N(x/4)=0, N=#O, --> even veel tetra als O, aantal O_Vac=#O*((x/2)/2), eerste deling door twee wegens 1 vac per 2 La, tweede deling door twee wegen 2x zoveel O als cationen.
As a result, the attraction between Ce cations and O-vacancies can thus be seen as the driving force for order in the \LCO\ system. Such a link between the ordering of the cation and anion sub-latices is also found for other pyrochlores.\cite{Crocombette:NIMPRSB2007} The resulting geometry at $50$\% Ce substituted by La, is then expected to be a highly ordered structure containing Ce tetrahedra surrounding the O-vacancies. This  makes the pyrochlore geometry a very likely candidate for the \LCO\ geometry.

%%%%%%%%%%%%%%%%%%%%%%%%%%%%%%%%%%%%%%%%%%%%%%%%%%%%%%%%%%%%%%%%%%%%%%%%%%%%%%%%
%%%%%%%%%%%%%%%%%%%%Subsection:Results:Geometry%%%%%%%%%%%%%%%%%%%%%%%%%%%%%%%%%
%%%%%%%%%%%%%%%%%%%%%%%%%%%%%%%%%%%%%%%%%%%%%%%%%%%%%%%%%%%%%%%%%%%%%%%%%%%%%%%%
\subsection{Lattice parameter}
%\begin{table}[!tb]
%\caption{The same as Table \ref{table:LCO_GeomPBE}, but for the PBE+U results.}\label{table:LCO_GeomPBEU}
%\begin{ruledtabular}
%\begin{tabular}{l|cccc}
%& V & a & \multicolumn{2}{c}{$\Delta$a} \\
%&   &   &  vs CeO$_2$ & vs NoVac \\
%&  (\%)&  (\AA)&  (\%)&  (\%)\\
%\hline
%c111 L$1_0$ NoVac & 7.945 & 5.609 & 2.581 & $-$ \\
%c111 L$1_0$ 2Ce2La & -1.631 & 5.579 & 2.021 & -0.547 \\
%p222 L$1_0$ NoVac  & 7.947 & 5.609 & 2.582 & $-$ \\
%p222 L$1_0$ 2Ce2La & -1.725 & 5.577 & 1.989 & -0.578 \\
%\hline
%LZO NoVac  & 8.479 & 5.618 & 2.750 & $-$ \\
%LZO 2Ce2La & -1.533 & 5.590 & 2.222 & -0.514 \\
%LZO 4Ce    & -0.228 & 5.614 & 2.672 & -0.076 \\
%LZO 4La    & -3.793 & 5.547 & 1.434 & -1.281 \\
%\hline
%L111 NoVac   & 8.639 & 5.621 & 2.800 & $-$ \\
%L111 3Ce1La a  & -1.045 & 5.602 & 2.441 & -0.350 \\
%%L111 3Ce1La b & -0.889 & 5.605 & 2.495 & -0.297 \\
%%L111 3Ce1La c & -0.459 & 5.613 & 2.643 & -0.153 \\
%L111 1Ce3La  & -2.774 & 5.569 & 1.841 & -0.933 \\
%\hline
%SQS Novac & 9.173 & 5.630 & 2.969 & $-$ \\
%SQS   & -1.285 & 5.606 & 2.526 & -0.430 \\
%\end{tabular}
%\end{ruledtabular}
%\end{table}

\indent In addition to the energies, we also investigated the cell volume and lattice parameter. Table~\ref{table:LCO_GeomPBE} shows the expansion and contraction of the system compared to the pure CeO$_2$ fluorite structure. Most configurations studied show small distortions of the lattice vectors compared to the cubic fluorite lattice. Only the LZO NV, 4Ce, and 4La configurations retained a cubic lattice.\\
\indent Because the distortions of the different system lattices are all slightly different, it is impossible to extract a (consistent) lattice parameter from their basis vectors. However, since the distortions are relatively small, less than $4^{\circ}$ for the lattice vector angles, and less than $3.5$\% for the lattice vector lengths, we instead choose to use a fictitious lattice parameter, which is determined as the edge of a cube containing a single formula unit, with the same mass density as the actual crystal.\\
\indent Comparing the results in Table \ref{table:LCO_GeomPBE} shows the Coulomb corrections to have only limited influence on the geometry. Compared to the CeO$_2$ fluorite system, the \LCO\ systems without vacancies are roughly $8$\% larger in volume. The introduction of O-vacancies causes a small contraction of the lattice parameter of a few tenth of a percent compared to the NV systems. This contraction appears to be correlated to the amount of Ce atoms present in the tetrahedral vacancy surrounding; more Ce results in less contraction. This behavior can be understood as a consequence of the increase in ionic radius known for the \CeIV $ \rightarrow$ \CeIII\ transition.\cite{Shannon:ACSA1976,Shannon:table}\\
\indent It is also interesting to note that for the L111 3Ce1La cases the lattice parameter seems to increase with decreasing O-vacancy distance, in contrast to what is intuitively expected.\\
\indent Looking at the three most stable configurations with regard to \Evf\ of the previous section (LZO 4Ce, L111 3Ce1La B and C), we find that these are also the three systems having the largest lattice parameter. Their lattice parameter is $5.61$ \AA, which is $\sim2.6\%$ larger than the normal CeO$_2$ fluorite structure. This is in good agreement with the experimentally observed value of $5.53-5.61$ \AA\ for the \LCO\ lattice parameter.\cite{BrisseF:CJChem1967,CaoXueqiang:AdvMat2003,BaeJongSung:JoECS2004,BelliereV:JPhysChemB2006,MorrisBC:JMatChem1993}
It also shows good agreement with the experimentally observed relative expansion of $\sim2.8$\% by Ryan \textit{et al.} and Belli\`{e}re \textit{et al.}\cite{RyanKM:JPCondMat2003,BelliereV:JPhysChemB2006}
In addition, the pyrochlore structure LZO 4Ce is the only  one of these three structures which retained a perfect cubic lattice, as expected from experiment for \LCO.\cite{BrisseF:CJChem1967}\\
\subsection{X-Ray diffraction}
\begin{figure}[tb!]
  \includegraphics[width=8cm,keepaspectratio=true]{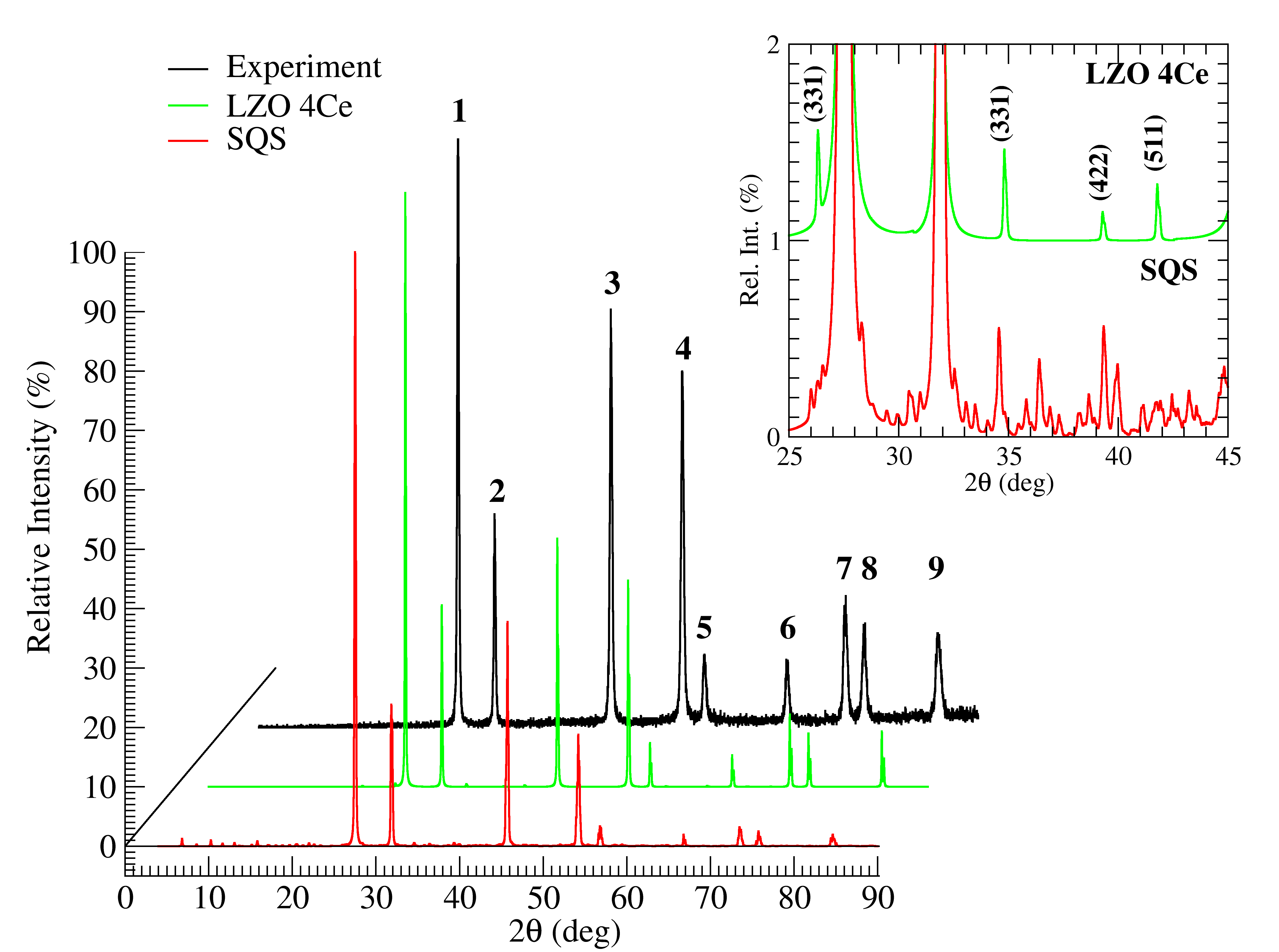}\\
  \caption{(Color online) Comparison of calculated XRD images for the disordered fluorite (SQS, red curve at  the front) and pyrochlore (LZO 4Ce, green curve in the middle) geometries to the experimental \LCO\ spectrum (black curve at the back).\cite{Vyshnavi:XXX} The relaxed PBE geometries are used to calculate the XRD spectra. The nine peaks with the highest intensity are indicated with indexes $1$ through $9$. All spectra are normalized with regard to the peaks with index $1$.
  %For clarity the experimental curve was shifted 2$^{\circ}$ and 10\%, and the LZO 4Ce curve was shifted 1$^{\circ}$ and 5\%.
  The inset shows a section of the spectra containing the pyrochlore (311), (331), and (511) peaks. The LZO 4Ce curve was shifted vertically by $1$\% for clarity.
  }\label{fig:calcXRDvsExpXRD}
\end{figure}
\begin{figure}[tb!]
  \includegraphics[width=8cm,keepaspectratio=true]{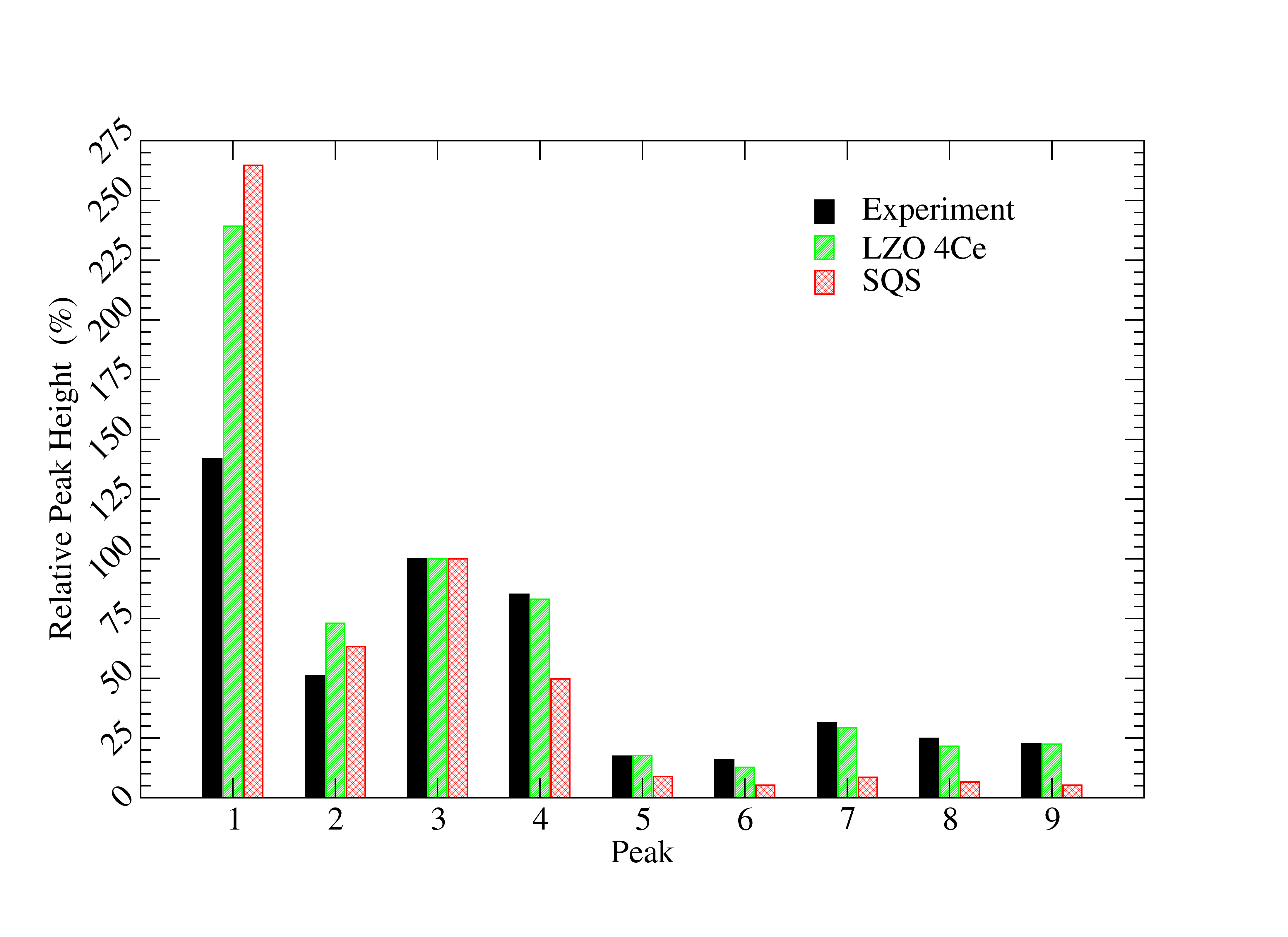}\\
  \caption{(Color online) Comparison of the nine high intensity peaks present in the spectra shown in Fig.~\ref{fig:calcXRDvsExpXRD}. Scaling was done with regard to the third peak for each spectrum.}\label{fig:calcXRDvsExpXRDbars}
\end{figure}
%\subsection*{XRD spectra}
\indent An important argument in the discussion on the structure of \LCO\ is its XRD spectrum. The differences between the \CO2\ fluorite and a prototypical pyrochlore spectrum are very small and mainly related to the presence of the low intensity peaks of the pyrochlore ($311$), ($331$), and ($511$) reflections.\cite{BaeJongSung:JoECS2004,CaoXueqiang:AdvMat2003,YamamuraH:SSI2007} The strong similarity between the \CO2\ and pyrochlore spectra is due to the nearly equal X-ray scattering factors of La and Ce.\cite{BrisseF:CJChem1967} As a result very sensitive XRD experiments are required if one is to observe the low intensity peaks indicative of the pyrochlore structure of \LCO.
\indent To address this point in the discussion, we calculated XRD spectra for all structures studied. Figure~\ref{fig:calcXRDvsExpXRD} shows calculated spectra of the SQS (disordered fluorite) and LZO 4Ce (pyrochlore) geometries in comparison to an experimental powder XRD spectrum of \LCO.\cite{Vyshnavi:XXX} The experimental spectrum shows nine clear peaks that can be identified with the fluorite peaks of \CO2, no small secondary pyrochlore peaks are visible. This is in line with the XRD spectra of \LCO\ generally presented in literature. The two calculated spectra in Fig.~\ref{fig:calcXRDvsExpXRD} represent the spectra of the two possible structures for the \LCO\ structure. In both calculated spectra the nine experimental peaks are clearly visible and their position is still within $1^{\circ}$ of the experimental position for the high angle peaks. Furthermore, no additional peaks with intensities $>1$\% are visible in the calculated spectra.\\
\indent It is interesting to note that even if a pyrochlore geometry is used, the typical pyrochlore peaks appear to be lacking. However, closer examination of the numerical data shows these peaks to be present, albeit at a much lower intensity than would be expected from other pyrochlore systems such as for example \LZO. The (311), (331) and (511) peaks have an intensity of $0.5$\% or less (\textit{cf}.~inset Fig.~\ref{fig:calcXRDvsExpXRD}). This explains why these peaks are not generally observed in experiments, but only in case of long time step scans.\cite{BaeJongSung:JoECS2004} This might lead to an erroneous assignment of the disordered fluorite structure to \LCO. The SQS structure on the other hand presents a large number of very small secondary reflections. Though they might not be clearly distinguishable in experiments, they should introduce a noticeable additional background (\textit{cf}.~inset Fig.~\ref{fig:calcXRDvsExpXRD}). This should be especially clear in experiments where increasing La concentrations in \LCOg\ are investigated. However, such experiments do not mention such behavior.\cite{BaeJongSung:JoECS2004,RyanKM:JPCondMat2003}\\
\indent A thorough comparison of the nine high intensity peaks also favors the pyrochlore (LZO 4Ce) over the disordered fluorite (SQS) structure, as geometry for \LCO. The relative peak heights of the former match the experimental peak heights much closer, as is shown in Fig.~\ref{fig:calcXRDvsExpXRDbars}. We chose not to scale the spectra using the highest intensity peak since it appears to be overestimated for both geometries. By using a different peak as scaling reference the good correlation of the relative peak heights, with the experimental spectrum becomes much clearer.\\
     %Furthermore, the high angle peaks in the disordered fluorite case show a significant broadening which is also not observed in the experimental peaks.\\
\begin{figure}[tb!]
  \includegraphics[width=8cm,keepaspectratio=true]{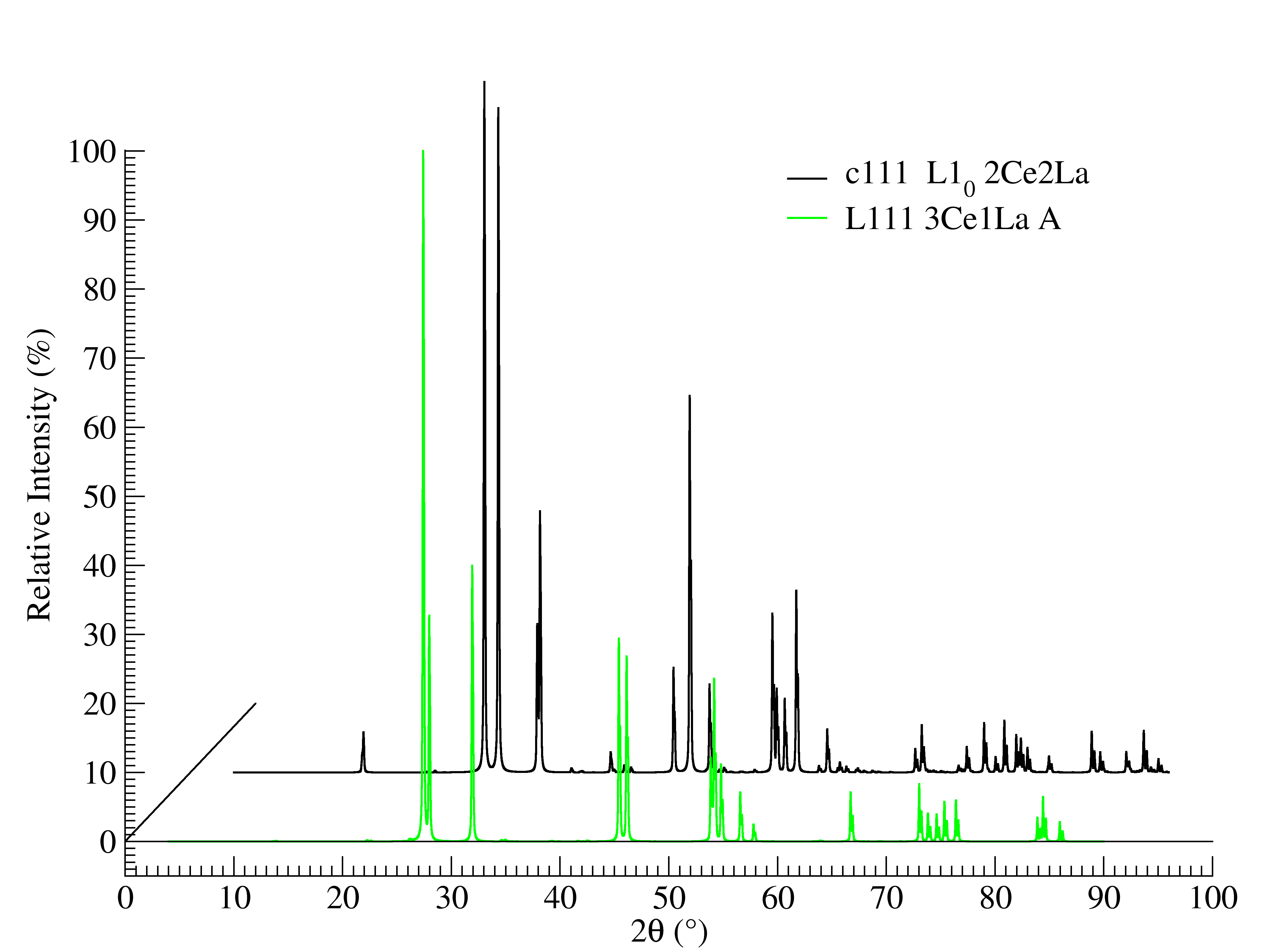}\\
  \caption{(Color online) Calculated XRD-spectra of the c$111$ L$1_0$ $2$Ce$2$La and L$111$ $3$Ce$1$La A structures. Multiple peak splitting is clearly visible for the high angle peaks.}\label{fig:calcXRDExtraSpectra}
\end{figure}
\indent As a final note, it is interesting to mention that the four high angle peaks, can also be used to exclude the other structures studied in this work. This is because of the broadening of those peaks due to multiple splitting. As a consequence, the two peaks between $70$ and $80^{\circ}$ merge into a near continuous set of peaks spread over a range of $5^{\circ}$ in case of the L111 structures, and nearly $10^{\circ}$ in case of the L$1_0$ structures. Two examples are given in Fig.~\ref{fig:calcXRDExtraSpectra}. No simple peaks, as present in the curves of Fig.~\ref{fig:calcXRDvsExpXRD} are observed for those structures.\\

\section{Conclusion}
\indent In this paper the geometry of \LCO\ was studied using \textit{ab initio} DFT calculations. Ordered and disordered structures were compared, modeling the pyrochlore and the disordered fluorite structures proposed for this system. Three aspects are taken into consideration: the lattice parameter, the enthalpy of formation and the XRD spectrum.\\
\indent The calculated lattice parameters and relative lattice expansion of both the pyrochlore and disordered fluorite geometry are found to be in good agreement with the experimentally observed values. Of these the pyrochlore structure retained a perfect cubic lattice, while the disordered fluorite structure showed a small distortion.\\
\indent Charge compensating O-vacancies are shown to play a crucial role in the stability of \LCO. If no O-vacancies are present (\textit{i.e.} La$_2$Ce$_2$O$_8$) the disordered fluorite structure is the most stable geometry. However, if O-vacancies are included in the system, then the pyrochlore structure becomes the most stable geometry. So, in contrast to what is generally assumed in literature, \LCO\ is predicted to have a pyrochlore rather than a disordered fluorite geometry.\\
\indent Total energy calculations have also shown a clear preference of the O-vacancies to be centered in Ce tetrahedra. This might be seen as a driving force in the formation of a pyrochlore geometry during \LCO\ formation. The formation of \LCO, however, is beyond the scope of this work, and is, as such, a topic for further research.\\
\indent The third aspect, the XRD spectra, show for both the disordered fluorite and pyrochlore geometry the nine distinct high intensity peaks observed for \LCO\ in experiment. Although the peak positions are in excellent agreement with the experimental positions, only the pyrochlore geometry presents relative peak intensities that are in good agreement with the experimental ones. Furthermore, the pyrochlore geometry also presents the typical pyrochlore reflection peaks, albeit with a lower intensity than expected from other pyrochlores.\\
\indent Finally, both formation energies and calculated XRD spectra show a highly ordered structure to be preferred over a disordered structure, as model for the \LCO\ system. Of the structures considered in this work, the pyrochlore geometry is clearly favorable over the disordered fluorite geometry.\\

%Special thanks etc
%----------------------------------------------------------------------
\section{Acknowledgement}
\indent We would like to thank Olivier Janssens and Vyshnavi Narayanan for the XRD measurements. The research was financially supported by FWO-Vlaanderen, project n$^{\circ}$ G. $0802.09$N. We also acknowledge the Research Board of the Ghent University. All calculations were carried out using the Stevin Supercomputer Infrastructure at Ghent University. XRD measurements were carried out under the Interuniversity Attraction Poles Programme IAP/VI-$17$ (INANOMAT) financed by the Belgian State, Federal Science Policy office.

%----------------------------------------------------------------------

%*************************************************************************************************************
% BIBLIOGRAPHY
%*************************************************************************************************************
% Put in \nocite{*} so all entries in the bibliography are included
%\nocite{*}
% This GATHER command is useful for when you want to use WinEdt's Gather functionality, i.e., type
% \cite{} and a popup box appears with all of your citations to choose from.  Leave the % on the next line.
% The commented way of writing Gather is the only correct way of doing it !!!
%GATHER{danny.bib}
%GATHER{LCO.bib}
\bibliography{danny,LCO}

\end{document}